\documentclass[sigconf,natbib=true]{acmart}

\AtBeginDocument{%
  \providecommand\BibTeX{{%
    \normalfont B\kern-0.5em{\scshape i\kern-0.25em b}\kern-0.8em\TeX}}}

\copyrightyear{2024}
\acmYear{2024}
\setcopyright{acmlicensed}\acmConference[SIGIR '24]{Proceedings of the 47th International ACM SIGIR Conference on Research and Development in Information Retrieval}{July 14--18, 2024}{Washington, DC, USA} \acmBooktitle{Proceedings of the 47th International ACM SIGIR Conference on Research and Development in Information Retrieval (SIGIR '24), July 14--18, 2024, Washington, DC, USA}
\acmDOI{10.1145/3626772.3657686} 
\acmISBN{979-8-4007-0431-4/24/07}


\acmConference[SIGIR'24]{47th International ACM SIGIR Conference on Research and Development in Information Retrieval}{July 14--18, 2024}{Washington, WDC}
\usepackage{algorithm}
\usepackage{algorithmicx}
\usepackage{algpseudocode}
\usepackage{multirow}
\usepackage{enumitem}
\usepackage{graphicx}
\usepackage{subfigure}
\usepackage{soul}

\usepackage{balance}

\usepackage{color, xspace}

\newcommand{\etal}{\emph{et al.}\xspace}
\newcommand{\eg}{\emph{e.g.,}\xspace}
\newcommand{\ie}{\emph{i.e.,}\xspace}
\newcommand{\etc}{\emph{etc.}\xspace}
\newcommand{\name}{M\textsuperscript{3}oE\xspace}

\usepackage{color}

\author{Zijian Zhang}
\affiliation{%
  \institution{Jilin University}
  \country{}
  }
\affiliation{%
  \institution{City University of Hong Kong}
  \state{Jilin}
   \country{China}
   }
\authornote{Zijian Zhang is also with the Key Laboratory of Symbolic Computation and Knowledge Engineering of Ministry of Education, Jilin University.}
\authornote{This work is done during Zijian's internship at Kuaishou Technology.}
\email{zhangzj2114@mails.jlu.edu.cn}

\author{Shuchang Liu}
\affiliation{%
  \institution{Kuaishou Technology}
  \state{Beijing}
  \country{China}
   }
\email{liushuchang@kuaishou.com}

\author{Jiaao Yu}
\affiliation{%
  \institution{Kuaishou Technology}
  \state{Beijing}
  \country{China}
   }
\email{yujiaao596@163.com}

\author{Qingpeng Cai}
\affiliation{%
  \institution{Kuaishou Technology}
  \state{Beijing}
  \country{China}
   }
\email{cqpcurry@gmail.com}

\author{Xiangyu Zhao}
\affiliation{%
  \institution{City University of Hong Kong}
  \state{Hong Kong}
  \country{China}
  }
\email{xianzhao@cityu.edu.hk}
\authornote{Corresponding authors.}

\author{Chunxu Zhang}
\affiliation{%
  \institution{Jilin University}
  \state{Jilin}
   \country{China}
  }
\email{cxzhang19@mails.jlu.edu.cn}
\authornotemark[3]

\author{Ziru Liu}
\affiliation{%
  \institution{City University of Hong Kong}
  \state{Hong Kong}
  \country{China}
  }
\email{ziruliu2-c@my.cityu.edu.hk}

\author{Qidong Liu}
\affiliation{%
  \institution{Xi'an Jiaotong University} 
   \country{}
  \institution{City University of Hong Kong}
  \state{Xi'an}
   \country{China}
}
\email{liuqidong@stu.xjtu.edu.cn}

\author{Hongwei Zhao}
\affiliation{%
  \institution{Jilin University}
  \state{Jilin}
   \country{China}
  }
\email{zhaohw@jlu.edu.cn}
\authornotemark[3]

\author{Lantao Hu}
\affiliation{%
  \institution{Kuaishou Technology}
  \state{Beijing}
  \country{China}
   }
\email{hulantao@gmail.com}

\author{Peng Jiang}
\affiliation{%
  \institution{Kuaishou Technology}
  \state{Beijing}
  \country{China}
   }
\email{jp2006@139.com}

\author{Kun Gai}
\affiliation{%
  \institution{Unaffiliated}
   \country{}
  \state{Beijing}
  \country{China}
   }
\email{gai.kun@qq.com}

\begin{document}

\renewcommand{\shortauthors}{Zijian Zhang, et al.}
\title{
M\textsuperscript{3}oE: Multi-Domain Multi-Task Mixture-of-Experts Recommendation Framework
}

\begin{abstract}
Multi-domain recommendation and multi-task recommendation have demonstrated their effectiveness in leveraging common information from different domains and objectives for comprehensive user modeling. Nonetheless, the practical recommendation usually faces multiple domains and tasks simultaneously, which cannot be well-addressed by current methods. To this end, we introduce \name, an adaptive Multi-domain Multi-task Mixture-of-Experts recommendation framework. \name integrates multi-domain information, maps knowledge across domains and tasks, and optimizes multiple objectives. We leverage three mixture-of-experts modules to learn common, domain-aspect, and task-aspect user preferences respectively to address the complex dependencies among multiple domains and tasks in a disentangled manner. Additionally, we design a two-level fusion mechanism for precise control over feature extraction and fusion across diverse domains and tasks. The framework's adaptability is further enhanced by applying AutoML technique, which allows dynamic structure optimization. To the best of the authors' knowledge, our \name is the first effort to solve multi-domain multi-task recommendation self-adaptively. Extensive experiments on two benchmark datasets against diverse baselines demonstrate \name's superior performance. The implementation code is available to ensure reproducibility\footnote{\url{https://github.com/Applied-Machine-Learning-Lab/M3oE}}.
\end{abstract}


\maketitle
\section{Introduction}
With the rapid growth of web services and user-generated data, recommendation services have devoted increasing attention to a more precise understanding of user preferences and content recommendation \cite{sharma2013survey,isinkaye2015recommendation, wang2023multi,jia2024erase}.
As a result of this trend, the research on the recommender system has gradually moved towards increasingly complex but more practical scenarios like multi-domain~\cite{star, ddtcdr, li2023hamur, jia2024d3, gao2023autotransfer, li2022gromov} and multi-task~\cite{shared_bottom, mmoe, ple, liu2023multi} problems.
{The multi-domain recommendation problem assumes that users may exhibit similar tastes and behaviors across domains or platforms, indicating the feasibility of auxiliary information transfer \cite{mdlsurvey}.
For example, a user Mike watched Sci-Fi movies on TV, and is now browsing movies on the tablet.
Then knowing Mike's movie-watching history on TV would help figure out the preferences of Sci-Fi, and benefit the recommendation on the tablet domain.
The same effects also exist in reverse. 
}
In the multi-task scenario, users may interact or consume each recommended item in various ways, which brings multiple perspectives to describe a user's preference \cite{forkmerg}. 
For instance, in a video-sharing platform, users can watch, like, and save videos.
In reality, different signals may not always be positively related, \eg a user watched a video does not necessarily mean that the user likes or will save the video.
A more considerate recommender should be able to model and employ these complicated relations between tasks during inference.

In this context, recommender system designs have recently witnessed significant progress in both scenarios.
{On one hand, the key question in the multi-domain recommendation scenario is how to accurately extract relevant information and efficiently transfer between different domains.
For example, DDTCDR \cite{ddtcdr} designs a latent orthogonal mapping to transfer embedding between two domains.}
STAR \cite{star} leverages factorized domain-shared and domain-specific adaptive networks with a star topology to address the domain characteristics and their commonalities. 
By introducing relevant knowledge from different domains, Multi-Domain Recommendation (MDR) can leverage the beneficial insights to enhance the user personalization modeling in each domain \cite{adasparse,causalint,wang2024diff,liu2024multifs,wang2023plate}.
On the other hand, the multi-task recommendation scenario set up learning tasks for each of the feedback signals (\eg click-through rate prediction, like rate prediction, and save rate prediction), and the key challenge is how to balance different tasks and collaboratively improve the overall performance.
For example, Shared Bottom \cite{shared_bottom} utilizes a shared bottom layer to extract cross-task information.
MMoE \cite{mmoe} designs common expert networks and task-specific towers for multi-task modeling.
By learning user engagement from diverse aspects, Multi-Task Recommendation (MTR) can exploit the shared information between tasks and benefit from a more comprehensive understanding of user behavior \cite{mmoe,adatt,forkmerg}.
Though the tasks may also negatively influence each other like the seesaw phenomenon~\cite{ple,adatask} and so do the domains~\cite{star,pepnet}, in general, collaborate multiple learning tasks or incorporating information from multiple domains provides a more complete view of user-item interactions and helps improves the generality of recommendation.



Despite the progress of existing MDR and MTR methods, the actual recommendation environment nowadays is usually multi-domain and multi-task at the same time~\cite{m2m}.
This involves domain-task interplay that brings new challenges for information transfer and objective balancing.
However, limited work except~\cite{m2m} explored the combination between MDR and MTR and there is still a gap before reaching a comprehensive Multi-Domain Multi-Task (MDMT) solution.
In our empirical study, we observe that merely including an MDR or MTR solution only achieves sub-optimal or imbalanced performance.
\begin{figure}[!t]
\setlength{\abovecaptionskip}{-2mm}
\setlength{\belowcaptionskip}{-6.5mm}
{\subfigure{\includegraphics[width=1\linewidth]{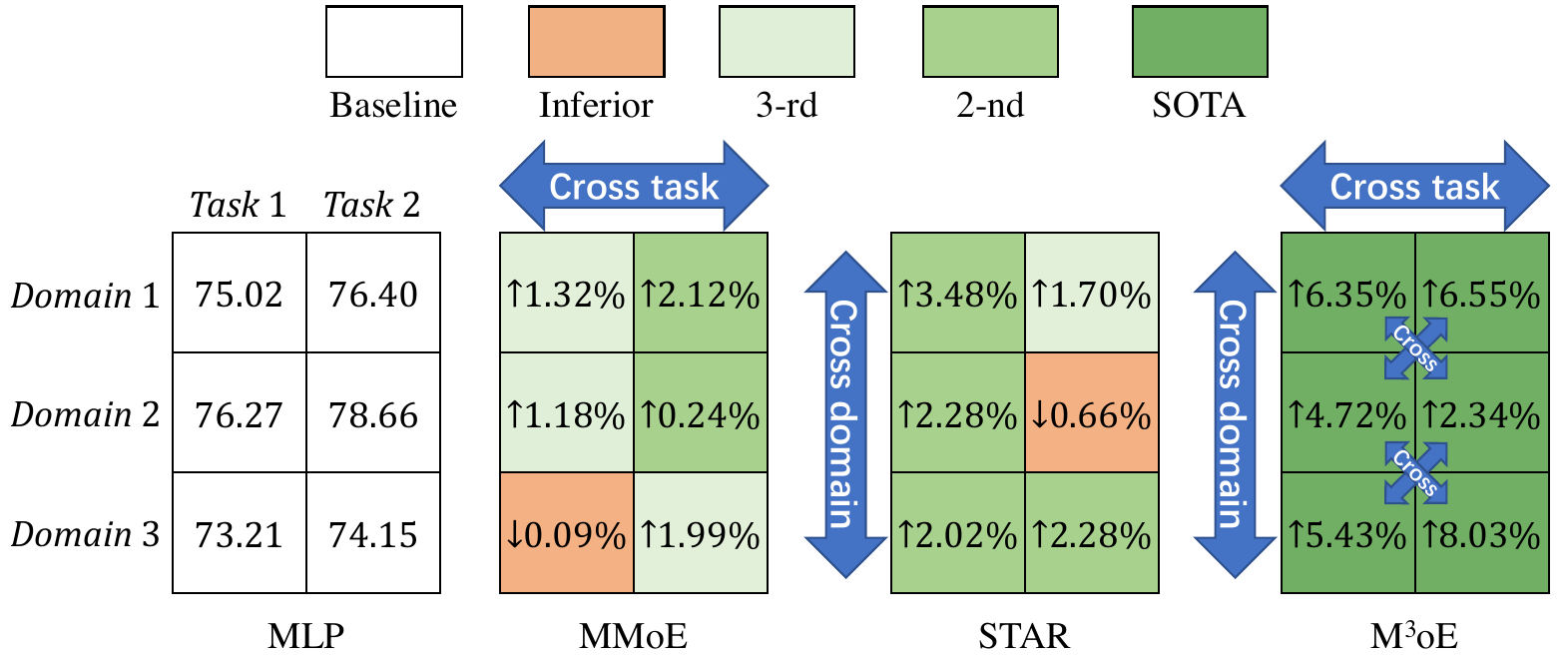}}}

    \caption{Multi-domain multi-task AUC comparisons on MovieLens. 
    We report the relative improvement of MMoE, STAR, and our \name, compared with single-domain single-task MLP baseline.
    The different colors indicate different performance ranks in the same domain and task.
    }
    \label{fig:pre}
\end{figure}
As an intuitive illustration of this phenomenon, Figure \ref{fig:pre} shows the ranking performance (\ie AUC) of representative methods of different types of existing solutions: 
MLP as single-domain single-task solution, 
MMoE \cite{mmoe} as MTR solution for each domain, and 
STAR \cite{star} as MDR solution for each task.
Though MMoE and STAR improve the overall performance over the single-domain single-task baseline, they may get worse in some special cases, as indicated by orange grids.
Besides, neither MMoE nor STAR is consistently better than the other for any single domain-task pair, indicating their sub-optimality in all domains and tasks.
In general, we find it really hard for an MDR or MTR solution to generalize well in the MDMT setting.

In analogy to the domain seesaw \cite{star,pepnet} and task seesaw \cite{ple,adatask} phenomena, we address this new challenge as MDMT seesaw and further describe it in two aspects:
\begin{itemize}[leftmargin=*]
    \item The same multi-domain information transfer method may not generalize to different tasks; and
    \item The same multi-task optimization balancing strategy may not generalize to different domains.
\end{itemize}
Again, using the video-sharing platform as an example: the domain seesaw addresses how to transfer a user' preferences from TV to tablet domain, and the task seesaw addresses how to balance user's behavior of watching and liking.
In contrast, the MDMT seesaw addresses problems like how to transfer a user's preference of watching on TV to augment user's preference of liking on tablet.
We argue that the key to this problem lies in how well we generalize the multi-domain multi-task knowledge transfer and integration mechanism, which has been long neglected in existing works.

In this paper, we propose a novel framework \name to jointly model multi-domain information extraction and multi-task inferences as a solution to the aforementioned MDMT seesaw challenge.
Specifically, it consists of a domain representation extraction layer that implements feature-level multi-domain knowledge transfer, a sophisticated multi-view expert learning layer that extracts and integrates information for specific domains and tasks, and an MDMT objective prediction layer that generates the separate output for each domain-task pair.
To better extract transferable information across domains for all tasks, we employ three essential types of experts in the middle expert learning layer to leverage shared common interests, domain-aspect user preferences, and task-aspect user preferences, respectively.
As we will discuss in Figure \ref{fig:vis}, this disentangled design helps capture different views of the input.
With this multi-view knowledge extracted, we use a flexible two-level fusion module to control the information aggregation for each domain and task: while the first level controls the integration between domains and between tasks, the second level controls the integration between shared experts, domain experts, and task experts.
Furthermore, to pursue the generality for different datasets and recommendation environments, we leverage AutoML to optimize the fusion weights adaptively.
Empirically, our proposed \name can effectively engage cross-domain cross-task knowledge transfer and integration.
As the illustrative example in Figure \ref{fig:pre}, \name (rightmost) achieves consistent improvement over separate MLP, multi-domain, and multi-task baselines.
We summarize our main contributions as follows:
\begin{itemize}[leftmargin=*]
    \item We identify the MDMT seesaw problem in a practical recommendation environment and point out the insufficient generality of sole multi-domain or multi-task solutions.
    \item To the best of our knowledge, our proposed framework \name is the first general framework to solve MDMT recommendation self-adaptively.
    \item Extensive experiments on two benchmark datasets demonstrate the superior performance of \name against state-of-the-art and in-depth analysis support its efficacy on knowledge disentanglement and integration.
\end{itemize}

    


\section{Methodology}



\subsection{Problem Definition}
{
\noindent\textit{\textbf{Multi-Domain Multi-Task Recommendation.}}
}
{
Let $\boldsymbol{\mathcal{U}}$ and $\boldsymbol{\mathcal{I}}$ represent the user set and item set.
And the MDMT recommendation problem aims to find a solution that simultaneously optimizes $T$ recommendation tasks on $D$ domains.
We assume that all users, items, and tasks may overlap across domains, but each sample is observed in a certain domain (\eg a user interacts with an item either on TV or tablet, not both).
As a result, for a sample in domain $d\in\{1,\dots,D\}$, we define a comprehensive feature input $\boldsymbol{x}_d$ that consists of user, item, and context information.
Similarly, each sample also observes target labels in $T$ tasks under domain $d$ and there is no label in other domains.
In our setting, we consider X-through rate prediction task (\eg click-through rate) for all tasks and each target label is a binary signal in {1,0} that defines whether the user provides positive feedback (\eg a click) or not.
Then, our research goal is to learn a X-through rate prediction function $\widehat{\boldsymbol{y}}_{d, t}=f^{d,t}(\boldsymbol{x}_d)$ for each task $t\in\{1,\dots,T\}$ and domain $d\in\{1,\dots,D\}$.

\subsection{Framework Overview}
As illustrated in Figure \ref{fig:framework},
we construct a general multi-domain multi-task framework, consisting of the domain representation extraction layer, multi-view expert learning layer, and MDMT objective prediction layer, from bottom to top.
In the multi-view expert learning layer, we design the shared expert module, domain expert module, and task expert module, to address the common information learning, domain-aspect user preferences, and task-aspect user preferences, respectively.
To achieve the model adaptivity, we assign a two-level fusion mechanism with automatically updated fusion weights.
After attaining the comprehensive representation, we infer the user preference for each objective individually.
The flexible model architecture and automatic fusion-weights optimization of our \name enable it to seamlessly adapt to various multi-domain and multi-task settings.


\begin{figure}[!t]
\setlength{\abovecaptionskip}{-2mm}
\setlength{\belowcaptionskip}{-6mm}
{\subfigure{\includegraphics[width=1\linewidth]{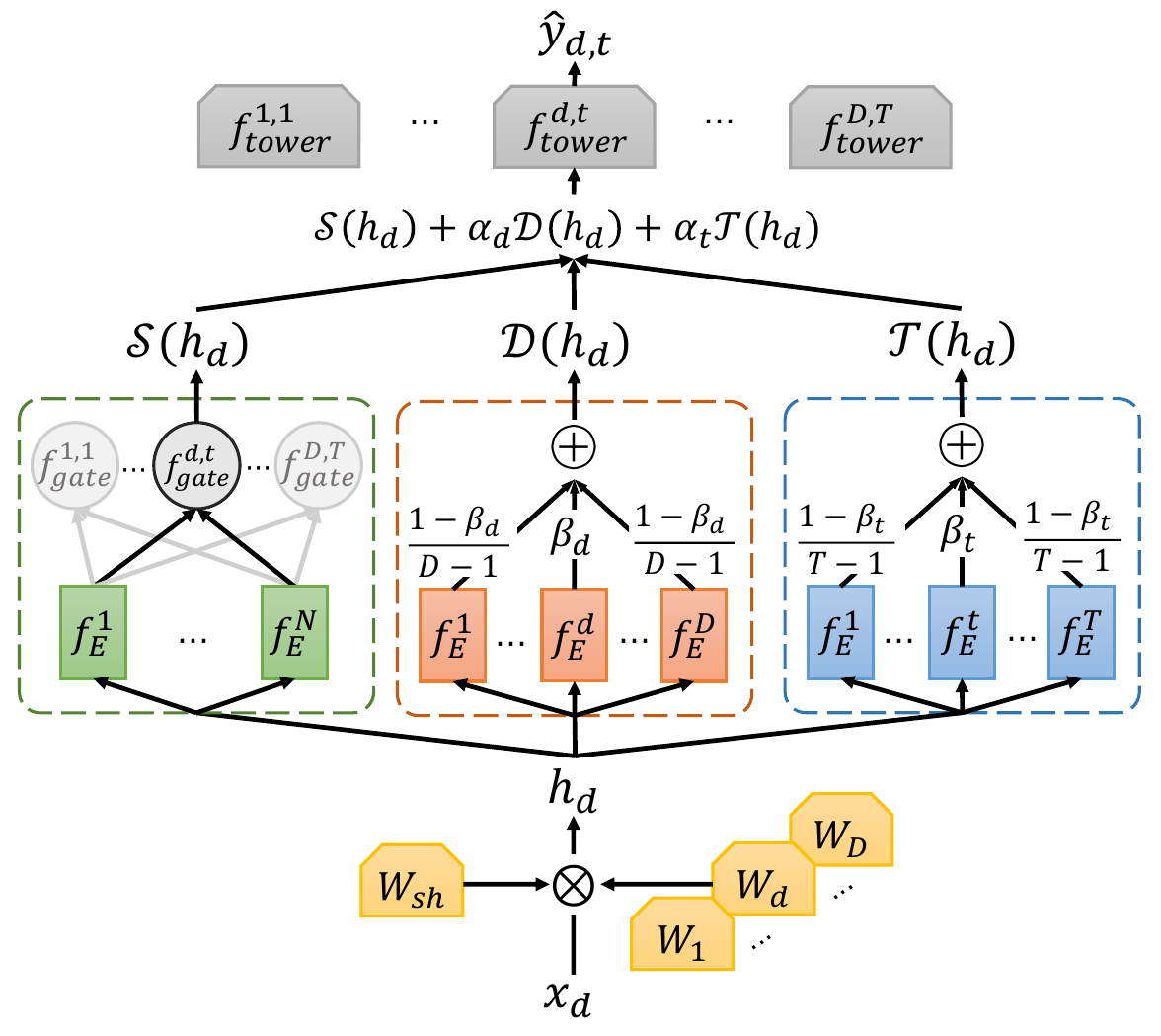}}}

    \caption{Framework of \name, focusing on domain $d$ and task $t$ for clarity.
    Within the multi-view layer, there are three modules arranged from left to right: shared expert module $\boldsymbol{\mathcal{S}}$, domain expert module $\boldsymbol{\mathcal{D}}$, and task expert module $\boldsymbol{\mathcal{T}}$.
    }
    \label{fig:framework}
\end{figure}

\subsection{Domain Representation Extraction Layer}
To unify the multi-domain representation and enhance the knowledge transfer, we introduce a domain-specific and common information integration structure for the input features.
Specifically, we incorporate fully connected layers to align attribute semantics and learn the representation of multiple domains \cite{star}. 
In this approach, we assign a domain-specific weight matrix $\boldsymbol{W}_d $ to handle domain-specific characteristics and utilize a shared weight matrix $\boldsymbol{W}_\mathrm{sh}$ that processes data from all domains to capture shared patterns.
To effectively combine the shared and domain-specific pattern, we perform an element-wise multiplication $\widehat{\boldsymbol{W}}_d = \boldsymbol{W}_d \otimes \boldsymbol{W}_\mathrm{sh}$, followed by the input vector processing.
The domain representation extraction operation $f_\mathrm{DR}$ on domain data $\boldsymbol{x}_d$ can be formulated as follows,
\vspace{-1mm}
\begin{equation}\label{Equ:star_w}
    f_\mathrm{DR}(\boldsymbol{x}_d) = \widehat{\boldsymbol{W}}_d\boldsymbol{x}_d + \boldsymbol{b}_d + \boldsymbol{b}_\mathrm{sh}
\end{equation}
\noindent where $\boldsymbol{b}_d$ and $\boldsymbol{b}_{sh}$ are the domain-specific and common bias.

After capturing domain common and distinct information, we perform a linear operation parameterized by a weight matrix $\boldsymbol{W}_c $ and a bias $\boldsymbol{b}_c$ to map the representation of different domains into the same embedding space. To mitigate the introduction of noisy information from other domains and regulate the unified representation space, we also include a multi-layer neural network $f_\mathrm{DA}$ that engages domain-agnostic mapping.
Then, given the sample input $\boldsymbol{x}_d$ of domain $d$, we extract its domain representation $\boldsymbol{h}_d$ as follows, 
\begin{equation}\label{Equ:star_mlp}
    \boldsymbol{h}_d = \boldsymbol{W}_c f_\mathrm{DR}(\boldsymbol{x}_d) + \boldsymbol{b}_c + f_\mathrm{DA}(\boldsymbol{x}_d)
\end{equation}
Note that $\boldsymbol{W}_\mathrm{sh}$, $\boldsymbol{W}_c$, and $f_\mathrm{DA}$ learn the patterns from all domains.
This would help generate basic features in a unified representation space for later information extraction modules.

\subsection{Multi-View Expert Learning Layer}
With the preprocessed representation $\boldsymbol{h}_d$, we also need to find a way to further extract and integrate information for each domain and task.
Intuitively, except for the common patterns (\eg common user interests), domains and tasks may potentially have different views of the given information.
In order to extract useful multi-view patterns in a disentangled and comprehensive manner, our approach explicitly defines three types of expert networks: 
\begin{itemize}[leftmargin=*]
    \item The shared expert module $\boldsymbol{\mathcal{S}}$ focuses on learning common knowledge that spans across domains and tasks.
\end{itemize}
\begin{itemize}[leftmargin=*]
    \item The domain expert module $\boldsymbol{\mathcal{D}}$ captures each domain's unique characteristics. Each expert is associated with one domain and is shared across tasks. 
\end{itemize}
\begin{itemize}[leftmargin=*]
    \item The task expert module $\boldsymbol{\mathcal{T}}$ models task-specific characteristics. Each task expert is associated with one task and is shared across domains.
\end{itemize}
For all expert network output, we design a two-level information fusion mechanism that learns to integrate extracted features for specific domains and tasks.
The first level learns to aggregate information between shared experts, between domains or between tasks, while the second level learns to aggregate information between three types of experts.
Without loss of generality, in the following sections, we show how different experts extract and integrate information for domain $d$ and task $t$.

\subsubsection{\textbf{Shared Expert Module}}
To capture common patterns across different domains and tasks, we employ multiple expert networks \cite{mmoe} and all experts will process input from multiple domain, \ie $\{h_d\}^{D}_{d=1}$.
In our approach, we incorporate $N$ shared expert networks, each of which is assigned a linear operation with a ReLU activation function.
Additionally, we introduce layer normalization to enhance the stability of the module learning process and improve embedding generalization. 
Note that the normalization technique ensures that the outputs of the expert networks have consistent distributions, promoting better alignment of the shared information.
Mathematically, the expert network $f_E^e$ for any expert $e\in\{1,\dots,N\}$ generates the following output:
\begin{equation}\label{Equ:expert}
    f^e_{E}(\boldsymbol{h}_d) = {\rm ReLU} ({\rm LayerNorm} (\boldsymbol{W}_e \boldsymbol{h}_d + \boldsymbol{b}_e))
\end{equation}
\noindent where $\boldsymbol{W}_e$ and $\boldsymbol{b}_e$ are trainable weights and bias of $f_E^e$.

\noindent\textbf{Shared Information Fusion.}
After learning the common information with $N$ shared expert networks, we propose a gate fusion mechanism to integrate these embeddings as shared information.
Concretely, we introduce a linear layer $f_\mathrm{gate}(\cdot)$ that generates an $N$-dimensional softmax weights for the sample-wise weighted sum of shared expert networks output. 
We allocate $D\times T$  gate layers to corresponding $D\times T$  objectives, ensuring that each objective has its own dedicated gate layer.
As depicted in the left part of Figure \ref{fig:framework}, domain $d$ and task $t$'s inference feeds on its corresponding gate-weighted embedding.
Hence, we can generate the embedding maintaining the shared information across multiple domains and tasks $\boldsymbol{\mathcal{S}}(\boldsymbol{h}_d)$ as follows,
\begin{equation}\label{Equ:shared_expert}
    \boldsymbol{\mathcal{S}}(\boldsymbol{h}_d) = f_\mathrm{gate}(\boldsymbol{h}_d) f_{E} (\boldsymbol{h}_d)
\end{equation}
\noindent where expert $f_{E}$ belongs to the shared expert network set $\boldsymbol{\mathcal{S}}$. 
Notably, we utilize $D\times T$ gate layers to assign a unique shared information fusion layer $f_\mathrm{gate}(\cdot)$  to each inference objective.

\subsubsection{\textbf{Domain Expert Module}}
The shared experts capture the common patterns, but each domain may have its own view about the input information and various tasks.
Following this notion, the domain expert module $\boldsymbol{\mathcal{D}}$ establish $D$ expert networks and we denoted ${\{f_{E}^{d}(\cdot)}\}$ for each domain's expert, \ie $\boldsymbol{\mathcal{D}}:=$ ${\{f_{E}^{d}(\cdot)}\}_{{d=1}}^D$, to address domain-specific information explicitly.
The network architecture is similar to that of the shared experts and has separate learnable parameters for different domains.

\noindent\textbf{Multi-Domain Disentanglement and Fusion.}
Different from shared experts that implicitly learn disentangled views, the domain experts are explicitly defined for input from different domains.
Intuitively, for samples $\boldsymbol{h}_d$ from domain $d$, we should focus on the output from the corresponding expert of domain $d$ and obtain the representation $f_E^d(\boldsymbol{h}_d)$.
Besides, $\boldsymbol{h}_d$ is in the same representation space with other domains' input, so using other domain experts to process $\boldsymbol{h}_d$ also becomes valid and potentially augments the information extraction.
Specifically, we feed $\boldsymbol{h}_d$ to the expert networks of other domains and denoted the obtained representations as $f_E^{\widetilde{d}}(\boldsymbol{h}_{d})$, where $\widetilde{d} \in \{1,\cdots,D|\widetilde{d}\neq d\}$.
In order to control the incorporation of information from other domain experts' representations, we set an affine balance weight $\beta_d \in (0, 1)$ that determines the balance between the current domain's representation and the information from other domain experts' descriptions. 
The multi-domain fusion embedding of domain $d$ can be calculated as follows,
\vspace{-2mm}
\begin{equation}\label{Equ:domain_f}
    \boldsymbol{\mathcal{D}}(\boldsymbol{h}_d) = \beta_d \cdot f_{E}^d (\boldsymbol{h}_d)+
    \frac{1-\beta_d}{D-1} \cdot \sum_{\widetilde{d}\neq d}^{D} f_{E}^{\widetilde{d}}(\boldsymbol{h}_d)
\end{equation}
where we assign equal weights $\frac{1-\beta_d}{D-1}$ to the representations from other domain expert networks to stabilize the augmented information and facilitate model training.
The sum of weights over all domain experts is 1, which guarantees that the weights maintain the desired balance.
In other words, we consider unbiased information integration for other domain experts' perspectives, while the expert of the corresponding domain $d$ receives focused attention through $\beta_d$.

\subsubsection{\textbf{Task Expert Module}}
To depict user interest from multiple tasks' aspects, we assign $T$ expert networks for the task expert module $\boldsymbol{\mathcal{T}}:=$ ${\{f_{E}^{t}(\cdot)}\}_{{t=1}}^{T}$. 
Similar to the domain expert module, each task is associated with a dedicated expert network, allowing explicit handling of task-specific information.

\noindent\textbf{Multi-Task Disentanglement and Fusion.}
Different from domain experts, the same input $\boldsymbol{h}_d$ is processed by all task experts and generates disentangled output.
Intuitively, each task should have its own focused view, so we adopt a similar biased fusion strategy where for the task $t$ we calculate the following,
\begin{equation}\label{Equ:task_f}
    \boldsymbol{\mathcal{T}}(\boldsymbol{h}_d) = \beta_t \cdot f_{E}^t (\boldsymbol{h}_d)+
    \frac{1-\beta_t}{T-1} \cdot \sum_{\widetilde{t}\neq t}^{T} f_{E}^{\widetilde{t}}(\boldsymbol{h}_d)
\end{equation}
\noindent where $f_{E}^t(\cdot)$ is expert network of task $t$, $f_{E}^{\widetilde{t}}(\cdot)$ means expert networks of other tasks.
Similar to the domain expert module, we assign unbiased weights to other task expert networks to ensure training stability and assign $\beta_t$ for the focused task. 
From the task viewpoint, it is important to note that the data samples from all $D$ domains share the same task expert network. 
This enables the task expert module to integrate task-related domain-agnostic information, enhancing the understanding of user preferences within the context of each task.

\subsubsection{\textbf{Multi-View Representation Balancing}}
Based on the shared expert module, we obtain common patterns $\boldsymbol{\mathcal{S}}(\boldsymbol{h}_d)$ across multiple domains and multiple tasks.
In addition, through the domain expert module and task expert module, we obtain multi-domain information $\boldsymbol{\mathcal{D}}(\boldsymbol{h}_d)$, and multi-task information $\boldsymbol{\mathcal{T}}(\boldsymbol{h}_d)$.
As we discussed earlier, performance in certain domains and tasks can benefit from multi-domain modeling and multi-task modeling to varying degrees.
At the same time, common knowledge between domains and tasks is also crucial in user preference modeling.
Therefore, we propose to balance these three components to achieve the final representation for downstream inference.
In specific, 
we allocate two weights $\alpha_d \in (0, 1)$ and $\alpha_t \in (0, 1)$ to balance the contribution of each component, achieving the comprehensive representation $\overline{\boldsymbol{h}}_d$ as follows,
\begin{equation}\label{Equ:multi_view_fusion}
    \overline{\boldsymbol{h}}_d = \boldsymbol{\mathcal{S}}(\boldsymbol{h}_d) + \alpha_d \cdot \boldsymbol{\mathcal{T}}(\boldsymbol{h}_d) + \alpha_t \cdot \boldsymbol{\mathcal{D}}(\boldsymbol{h}_d)
\end{equation}

By adjusting the values of $\alpha_d$ and $\alpha_t$, we can flexibly balance the contributions of the domain-specific, task-specific, and common knowledge components, enabling an adjustable representation that effectively captures the user preferences.

\noindent \textbf{Discussion of Two-Level Multi-domain Multi-task Fusion.}
In this paragraph, we take a close view of the elements of the embedding $\overline{\boldsymbol{h}}_d$.
In the two-level multi-domain multi-task fusion mechanism, 
the balance weight $\beta_t$ controls the trade-off between the domain-specific expert network and the experts from other domains. 
When $\beta_d \in (0.5, 1)$, the domain-specific expert network contributes more significantly compared to other domain experts. 
As $\beta_d$ approaches 1, it represents a scenario where only the domain-specific expert network is considered without incorporating information from other domains. 
Conversely, $\beta_d \in (0, 0.5)$ indicates that knowledge transfer from other domain expert networks contributes more than the domain-specific expert network.

Furthermore, $\alpha_d$ controls the overall weight of the domain expert module for generating domain representation $\overline{\boldsymbol{h}}_d$.
Higher values of $\alpha_d$ signify that the domain-aspect patterns play a more significant role compared to the common knowledge and task-specific perspectives. 
It also further recognizes the efficacy of combination inside the domain expert module.
Conversely, lower values of $\alpha_d$ indicate a stronger reliance on the other two components during training.
Based on $\alpha_d$ and $\beta_d$, we achieve precise control of factorized contribution from specific and shared domain experts.

By appropriately weighting and combining these components, we can leverage the advantages of multi-domain modeling, multi-task modeling, and shared knowledge to create a comprehensive and representative embedding that captures user preferences. 
The concrete balancing and fusion scheme will depend on the concrete requirements and characteristics of the application scenario.

\subsection{MDMT Objective Prediction Layer}
\label{subsec:prediction}
To infer the user preference across different tasks and domains, we introduce multi-domain multi-task objective prediction layer in this subsection.
The prediction process for a specific domain and task may depend on the diverse ingredients of domain representation differently. 
To accommodate the diverse numerical spaces associated with different domains and tasks, we assign individual MLP prediction towers, denoted as $f_{tower}^{d, t}(\cdot)$, for each unique domain and task combination represented as $(d, t)$.

The embeddings generated by the domain expert module and task expert module are fed to the corresponding prediction tower, respectively.
The separate inference modules allow for fine-grained control over the domain and task expert networks.
\begin{equation}\label{Equ:f_tower}
     f_{tower}^{d, t}(\overline{\boldsymbol{h}}_d)= \boldsymbol{W}_{d, t}^2{\rm ReLU}(\boldsymbol{W}_{d, t}^1 \overline{\boldsymbol{h}}_d + \boldsymbol{b}_{d, t}^1) + \boldsymbol{b}_{d, t}^2
\end{equation}
\noindent where $\boldsymbol{W}_{d, t}^2, \boldsymbol{b}_{d, t}^2, \boldsymbol{W}_{d, t}^1,$ and $ \boldsymbol{b}_{d, t}^1$ are weights and biases of two layers.

The model inferences of user preferences on samples from domain $d$ and task $t$ are calculated as,
\begin{equation}\label{Equ:f_tower}
     \widehat{\boldsymbol{y}}_{d, t} = {\rm Sigmoid}(f_{tower}^{d, t}(\overline{\boldsymbol{h}}_d))
\end{equation}
For binary classification tasks, we pick the Sigmoid function as the activation function of the prediction tower. 

\subsection{Optimization by AutoML}
The existing multi-domain and multi-task framework depends on manually designed architectures, which suffer from poor generality on new data and tasks.
Automated Machine Learning (AutoML) \cite{darts} has demonstrated its advanced adaptivity and structural flexibility in allocating architectures and hyper-parameters.

Considering the complex relationships among domains and tasks, we utilize AutoML to optimize the critical domain and task fusion weights $\alpha_d$, $\alpha_t$, $\beta_d$, and $\beta_t$.
Specifically, we generate each weight by a one-dimensional trainable tensor $\boldsymbol{e}_{w} \in \{\boldsymbol{e}_{\alpha_d}, \boldsymbol{e}_{\alpha_t}, \boldsymbol{e}_{\beta_d}, \boldsymbol{e}_{\beta_t}\}$ with a Sigmoid activation function respectively. Mathematically, the generation of $w \in \{\alpha_d, \alpha_t,$ $ \beta_d, \beta_t\}$ can be formulated as follows,
\begin{equation}\label{Equ:alpha}
     w = {\rm Sigmoid}(\boldsymbol{e}_{w})
\end{equation}

Then, we optimize these weights along with model training, \ie Bi-Level Optimization.
This end-to-end pipeline allows for the determination of optimal weights that correspond to the specific domains and tasks involved.

\subsubsection{Bi-Level Optimization}
Let $\boldsymbol{W}$ denote the model parameters of \name, $\boldsymbol{\alpha}:=\{\alpha_d, \alpha_t\}$, and $\boldsymbol{\beta}:=\{\beta_d, \beta_t\}$. 
This framework could be optimized with a Bi-Level Optimization, which optimizes the two parts of parameters alternatively.
We first update the model framework $\boldsymbol{W}$ for one epoch, then we calculate loss based on one batch data and optimize the weights $\boldsymbol{\alpha}$ and $\boldsymbol{\beta}$.
Notably, the update of $\boldsymbol{\alpha}$ and $\boldsymbol{\beta}$ are based on a mini-batch of training data, which takes trivial computational costs.
\begin{equation}
\label{Equ:automl}
\begin{aligned}
&\min _{{\alpha},{\beta}} \mathcal{L}\left(\boldsymbol{W}^{*}(\boldsymbol{\alpha},\boldsymbol{\beta}), \boldsymbol{\alpha},\boldsymbol{\beta}\right) \\
&\text { s.t. } \boldsymbol{W}^{*}(\boldsymbol{\alpha},\boldsymbol{\beta})=\arg \min _{\boldsymbol{W}} \mathcal{L}\left(\boldsymbol{W}, \boldsymbol{\alpha},\boldsymbol{\beta}\right)
\end{aligned}
\end{equation}

We select Binary Cross Entropy as the loss function,
\begin{equation}
\label{Equ:obj}
\mathcal{L} = \sum_d^{\boldsymbol{\mathcal{|D|}}}\sum_t^{\boldsymbol{\mathcal{|T|}}}{\rm BCE}(\widehat{\boldsymbol{y}}_{d, t}, \boldsymbol{y}_{d, t})
\end{equation}

\begin{table}[!t]
\renewcommand{\arraystretch}{1.1}
\centering
\setlength{\abovecaptionskip}{0.1mm}
\caption{Dataset statistics. 
}
\label{dataset}
\resizebox{0.5\textwidth}{!}{
\begin{tabular}{ccccccc}
\hline
\textbf{Dataset} & \multicolumn{3}{c}{\textbf{MovieLens}} &  \multicolumn{3}{c}{\textbf{KuaiRand-Pure}} \\
\cmidrule(lr){2-4}\cmidrule(lr){5-7}
\textbf{Domain} & 1 & 2 & 3 & 1 & 2 & 3 \\
\hline
\#Users & 1,325 & 2,096 & 2,619 & 15,398 & 27,049 & 11,809 \\
\#Items & 3,429 & 3,508 & 3,595 & 6,233 & 7,580 & 4,633 \\
\#Instances & 210,747 & 395,556 & 393,906 & 178,087 & 2,236,414 & 93,165 \\
Percentage & 21.07\% & 39.55\% & 39.38\% & 7.10\% & 89.18\% & 3.72\% \\
\hline
\end{tabular}}
\vspace{-3mm}
\end{table}

\section{Experiment}

In this section, we demonstrate the experiment results, including comparison with diverse advanced baselines, visualization of the disentangled and fusion of multi-domain multi-task user preference, the components contribution verification, and key hyper-parameters impact analysis.

\subsection{Dataset}
We evaluate the efficacy of \name on two public benchmark recommendation datasets, \ie MovieLens-1M and KuaiRand-Pure. The datasets' statistics are shown in Table \ref{dataset}. The split ratio of training, validation, and test is 8:1:1.

\begin{table*}[t!]
\renewcommand{\arraystretch}{1.2}
\centering
\setlength{\abovecaptionskip}{1mm}
\caption{Overall experiment results 
with unit of $\times 10^{-2}$.
``d1, t1'' means the result on domain 1 and task 1.
{``(a)'' means training on each domain and each task separately. 
``(b)'' means training multi-task models on each domain respectively. 
``(c)'' means training multi-domain models on each task respectively. 
``(d)'' means training on multi-domain multi-task setting. 
Best performances are bold, and the next best are underlined. 
``\textbf{{\Large *}}'' indicates the statistically significant improvements (\ie two-sided t-test with $p<0.05$) over the best baseline. 
All the results are the average of 5 repetitive runs using different random seeds.
}}
\label{tab:auc}
\scalebox{0.75}{
\begin{tabular}{clcccccccccccc||cccc}
\toprule
\multirow{2}{*}{} & \multirow{3}{*}{\textbf{Dataset}} & \multicolumn{12}{c||}{AUC for Each Domain and Task} & \multicolumn{4}{c}{Overall Performance} \\
\cmidrule(lr){3-14}\cmidrule(lr){15-18}
& & \multicolumn{6}{c}{MovieLens} & \multicolumn{6}{c||}{KuaiRand-Pure} & \multicolumn{2}{c}{MovieLens} & \multicolumn{2}{c}{KuaiRand-Pure} \\
\cmidrule(lr){3-8}\cmidrule(lr){9-14}\cmidrule(lr){15-16}\cmidrule(lr){17-18}
 & & \textbf{d1, t1} & \textbf{d1, t2} & \textbf{d2, t1} & \textbf{d2, t2} & \textbf{d3, t1} & \textbf{d3, t2} & \textbf{d1, t1} & \textbf{d1, t2} & \textbf{d2, t1} & \textbf{d2, t2} & \textbf{d3, t1} & \textbf{d3, t2} & \textbf{AUC $\uparrow$} & \textbf{Logloss $\downarrow$} & \textbf{AUC $\uparrow$} & \textbf{Logloss $\downarrow$} \\
\hline
(a) & MLP & 75.02 & 76.40 & 76.27 & 78.66 & 73.21 & 74.15 & 58.42 & 51.50 & \textbf{69.28} & 70.07 & 61.82 & 61.22 & 75.62 & 51.51 & 62.05 & 44.68 \\
\hline
\multirow{5}{*}{(b)} & ShBot-MTL & 75.36 & 76.25 & 75.90 & 78.40 & 72.51 & 73.30 & 61.33 & 56.13 & 68.62 & \underline{70.67} & 63.99 & 63.81 & 75.28 & 56.39 & 64.09 & 53.37 \\
 & PLE-MTL & 75.42 & 76.23 & 75.87 & 78.32 & 72.73 & 72.49 & 58.56 & 53.48 & 68.08 & 69.39 & 63.63 & 63.32 & 75.18 & 56.49 & 62.74 & 54.05 \\
 & MMOE-MTL & 75.35 & 76.96 & 76.58 & \underline{78.73} & 73.19 & 74.63 & 62.03 & 60.19 & \underline{68.81} & \textbf{70.91} & 64.08 & 64.32 & 75.90 & 51.27 & 65.05 & 44.07 \\
 & AdaTT & 72.54 & 74.24 & 74.12 & 73.01 & 69.75 & 70.25 & 61.92 & 60.59 & 64.86 & 67.28 & 61.40 & 60.65 & 72.32 & 58.72 & 62.78 & 44.41 \\
 & AdaTT-sp & 72.49 & 73.51 & 76.04 & 77.59 & 73.04 & 74.26 & 60.01 & 58.28 & 65.75 & 68.06 & 61.15 & 60.35 & 74.49 & 54.65 & 62.27 & 48.80 \\
\hline
\multirow{4}{*}{(c)} & ShBot-MDL & 75.70 & 76.50 & 76.58 & 78.10 & 73.88 & 74.76 & 64.29 & {63.36} & 65.83 & 67.22 & \underline{65.88} & \underline{65.47} & 75.92 & 51.53 & {65.34} & \underline{44.04} \\
 & MMOE-MDL & 75.77 & 76.50 & 76.71 & 78.37 & 73.92 & 75.16 & \underline{64.33} & 63.00 & 66.16 & 67.39 & 65.52 & 65.27 & 76.08 & 51.48 & 65.28 & 44.28 \\
 & PLE-MDL & 75.64 & 76.19 & 76.48 & 78.14 & 73.73 & 74.85 & 64.02 & 62.68 & 66.13 & 67.21 & 65.62 & 65.33 & 75.84 & 52.00 & 65.16 & 44.14 \\
 & STAR & 75.89 & 76.85 & \underline{76.87} & 78.47 & 73.68 & 74.70 & 64.12 & 62.96 & 64.91 & 66.32 & 65.73 & 64.28 & 76.08 & {51.11} & 64.72 & 44.68 \\
\hline
\multirow{4}{*}{(d)} & ShBot-MDMT & {75.94} & 77.25 & 76.56 & 78.63 & \underline{73.97} & 75.21 & 63.32& 63.55& 65.80& 68.04& 64.45& 64.89& 76.26 & \underline{50.96}& 65.01& 44.91\\
 & MMOE-MDMT & \underline{76.07} & \underline{77.27} & 76.86 & 78.67 & 73.83 & \underline{75.41} & 63.50& \underline{64.30}& 65.70& 67.88& 65.61& 66.10& \underline{76.35} & 51.02& \underline{65.52}& 44.20\\
 & PLE-MDMT & 75.83 & 76.95 & 76.04 & 78.48 & 73.67 & 75.09 & 61.86 & 61.88 & 63.57 & 66.09 & 64.45 & 63.47 & 75.68 & 51.90 & 64.07 & 44.67 \\
 & M2M & 74.12 & 75.35 & 75.08 & 77.13 & 72.11 & 73.78 & 61.29 & 62.25 & 60.44 & 63.15 & 61.75 & 61.57 & 74.59 & 54.14 & 61.74 & 44.55 \\
\hline
(ours) & \textbf{\name} & \textbf{76.61}{{\Large *}} & \textbf{78.13}{{\Large *}} & \textbf{77.51}{{\Large *}} & \textbf{79.33}{{\Large *}} & \textbf{74.47}{{\Large *}} & \textbf{76.09}{{\Large *}} & \textbf{64.85}{{\Large *}} & \textbf{65.89}{{\Large *}} & 66.03 & 68.31 & \textbf{66.21}{{\Large *}} & \textbf{66.91}{{\Large *}} & \textbf{77.02}{{\Large *}} & \textbf{50.71}{{\Large *}} & \textbf{66.37}{{\Large *}} & \textbf{43.76}{{\Large *}} \\
\bottomrule
\multicolumn{2}{c}{\textbf{RelaImpr$\uparrow$}} & 2.07\% & 3.15\% & 2.38\% & 2.09\% & 2.09\% & 2.68\% & 3.63\% & 11.12\% & -- & -- & 2.08\% & 9.31\% & 2.54\% & -- & 5.48\% & -- \\
\bottomrule
\end{tabular}}
\vspace{-3mm}
\end{table*}

\begin{itemize}[leftmargin=*]
    \item \textbf{MovieLens\footnote{https://grouplens.org/datasets/movielens/}} 
    The MovieLens dataset comprises 1 million ratings for around 3,900 movies and includes 7 user attributes and 2 item attributes. It encompasses a diverse range of data, including ratings and user information from various domains and tasks. It records user demographic information, such as gender, age, occupation \etc We use the feature ``age'' to separate dataset into 3 domains and infer ``click'' and ``like'' 2 tasks.
\end{itemize}

\begin{itemize}[leftmargin=*]
    \item \textbf{KuaiRand-Pure\footnote{https://kuairand.com/}}
    This dataset is collected from the short video platform Kuaishou\footnote{https://www.kuaishou.com/cn}, containing 21 user features, 12 video features, and 17 common features. We select ``tab'' representing interactions on different tabs to divide the dtaset into 3 domains and address ``click'' and ``long-view'' 2 tasks.

\end{itemize}

\subsection{Baseline}
To identify the performance comprehensively, we select advanced baselines for comparison.
Considering the diverse solutions to accomplish multi-domain and multi-task recommendation, we incorporate the MDR, MTR, and multi-domain multi-task versions of Shared Bottom, MMoE, and PLE.
In specific, we denote the suffix ``\textbf{-MTL}'' for the multi-task setting, which runs $D$ times individually for $D$ domains.
Denote ``\textbf{-MDL}'' for the multi-domain setting, which runs $T$ times individually for $T$ tasks.
Denote ``\textbf{-MDMT}'' for the multi-domain multi-task setting.

\begin{itemize}[leftmargin=*]
    \item \textbf{MLP}. 
    We utilize MLP architecture as a baseline performance for each specific task and domain. It includes a linear operation with a ReLU activation function and layer normalization. 
\end{itemize}

\begin{itemize}[leftmargin=*]
    \item \textbf{Shared Bottom} \cite{shared_bottom}. It is a representative multi-task recommendation model architecture, which enables the sharing of the bottom layer across tasks. 
    
\end{itemize}
\begin{itemize}[leftmargin=*]
    \item \textbf{MMoE} \cite{mmoe}. 
    The Multi-gate Mixture of Experts (MMoE) framework utilizes expert networks to address the capture of common information in multi-task learning. Each task has its dedicated prediction tower and a gate mechanism for information fusion. 
\end{itemize}

\begin{itemize}[leftmargin=*]
    \item \textbf{PLE} \cite{ple}. Progressive Layered Extraction (PLE) designs task-specific expert networks for each task, which enriches the information fusion mechanism with task-specific information. 
    
\end{itemize}

\begin{itemize}[leftmargin=*]
    \item \textbf{AdaTT} \cite{adatt}. 
    AdaTT is a multi-task learning approach that incorporates a branch consisting of task-specific experts, where the experts' outputs are linearly combined. 
    \textbf{AdaTT} refers to the overall approach that involves both task-specific and shared modules. \textbf{AdaTT-sp} refers to the version without the shared module.

\end{itemize}

\begin{itemize}[leftmargin=*]
    \item \textbf{STAR} \cite{star}. It is a multi-domain recommendation framework which designs star topology to integrate domain-specific and shared information effectively.
    
\end{itemize}

\begin{itemize}[leftmargin=*]
    \item \textbf{M2M} \cite{m2m}. M2M is a multi-scenario multi-task meta-learning approach, which utilizes a meta unit to leverage scenario knowledge, explicitly capturing inter-scenario correlations, along with a meta attention module and meta tower module to capture diverse inter-scenario correlations and enhance the representation of scenario-specific features, respectively.
    
\end{itemize}


\subsection{Experimental Setups}
We select the AUC and LogLoss to evaluate the performance and report all results as the average value of 5 repetitive runs using different random seeds. 
The embedding size is set to 16, and the learning rate is set to $1e-2$ for MovieLens and $3e-3$ for KuaiRand-Pure.
For the expert network, we employ a one-layer MLP, while the prediction layer consists of a two-layer MLP. 
In the shared expert module, we have 1 expert network for MovieLens, and 4 for KuaiRand-Pure. All expert networks across the three modules share the same structure.
To deploy on Movielens or KuaiRand-Pure with 3 domains and 2 tasks, our \name includes 3 expert networks in the domain expert module ($D=3$), and 2 expert networks in the task expert module ($T=2$). The prediction layer consists of 6 towers ($D\times T$) for the different combinations of domains and tasks.

\subsection{Overall Performance}
We compare \name with four lines of baselines: single task and domain, multi-task, multi-domain, and multi-domain multi-task recommendation.
We calculate Relative Improvement \cite{sarnet,jt_mdr}, denoted as \textbf{RelaImpr}, to show the improvement of \name over the best baseline.
From Table \ref{tab:auc}, we can safely make following conclusions:
\textbf{\textit{(1)}}
Multi-task methods (b) and multi-domain methods (c) generally outperform the single-domain single-task method MLP (a). This is attributed to the former methods' effective utilization of cross-task or cross-domain knowledge transfer, which offers an advantage in terms of input information compared to MLP, which solely relies on information from a single domain or single task.
However, an exception arises in domain 2 on KuaiRand-Pure, which contains a significantly larger number of samples, \ie ~90\% of the dataset, compared to the other two domains. As a result, cross-domain methods have to consider the performance balance across domains and yield inferior results.
\textbf{\textit{(2)}}
The multi-domain versions of the multi-task baselines, namely ShBot, PLE, and MMoE, consistently outperform their single-domain counterparts. 
This observation suggests that the performance gap between different domains is smaller compared to the gap between different tasks. It also indicates that leveraging cross-domain knowledge is generally easier than leveraging cross-task knowledge.
\textbf{\textit{(3)}}
Among the multi-domain methods (c), STAR demonstrates superior performance compared to the other approaches, highlighting the effectiveness of its domain representation learning. 
This result underscores the leading efficacy of STAR in capturing and leveraging domain-specific and common information for improved recommendation performance in multi-domain scenarios.
\textbf{\textit{(4)}}
In the multi-domain multi-task methods (d), M2M demonstrates its capability to leverage dependencies among domains and tasks, as it outperforms MLP on two tasks of domain 1 on the KuaiRand-Pure dataset. 
However, its performance falls short compared to the MDMT versions of MTR methods, \ie MMOE-MDMT, ShaBot-MDMT, and PLE-MDMT, highlighting its limited generality and capacity for joint modeling.
Furthermore, PLE-MDMT achieves better results on certain objectives than PLE-MTL and PLE-MDL. 
However, all three methods exhibit a seesaw phenomenon to some extent, and none of these methods consistently achieve a leading performance, emphasizing the need for further enhancements in their model capacity and optimization approaches.
\textbf{\textit{(5)}}
Our \name achieves consistent and almost all the best performance across all settings, which fully demonstrates its advanced ability to model multi-domain and multi-task jointly. 
On the MovieLens dataset, \name shows the dominant performance by surpassing all baselines on both average and specific objectives.
This achievement represents a significant breakthrough in comparison to other approaches, including multi-domain multi-task methods.
On the KuaiRand-Pure dataset, it surpasses all baselines on domain 1 and domain 3, except for domain 2 when compared with (a) and certain baselines in (b).
This can be attributed to domain 2 accounts for ~90\% of the samples, allowing methods solely trained on domain 2 to disregard balanced information sharing with other domains.
However, our \name's training process does not solely focus on domain 2 but aims to maintain a balanced performance across different domains, leading to remarkable improvements in domain 1 and domain 3, \eg 11.12\% improvement on domain 1 task 2 over the second best baseline.
Despite the highly imbalanced distribution of domain samples, our \name performs closely to the multi-task methods trained on a single domain. 
It is worth noting that \name achieves the best average AUC, \ie 5.48\% relative improvement over the best baseline, which represents the globally optimal solution. 
This highlights the effectiveness of our \name in achieving high-quality predictions across all domains and tasks simultaneously.

Hence, our \name well-addresses the multi-domain multi-task recommendation.
Its disentangled common knowledge, domain-aspect user preference, and task-aspect user preference help understand user personalization comprehensively.
Besides, the effective two-level fusion mechanism with adaptive weights precisely controls each factor's contribution to the specific objective.
Its self-adaptive optimization of fusion weights empowers the generality without extra human effort.
    
\vspace{-2mm}
\subsection{Visualization}
\begin{figure}[!t]
{\subfigure{\includegraphics[width=0.49\linewidth]{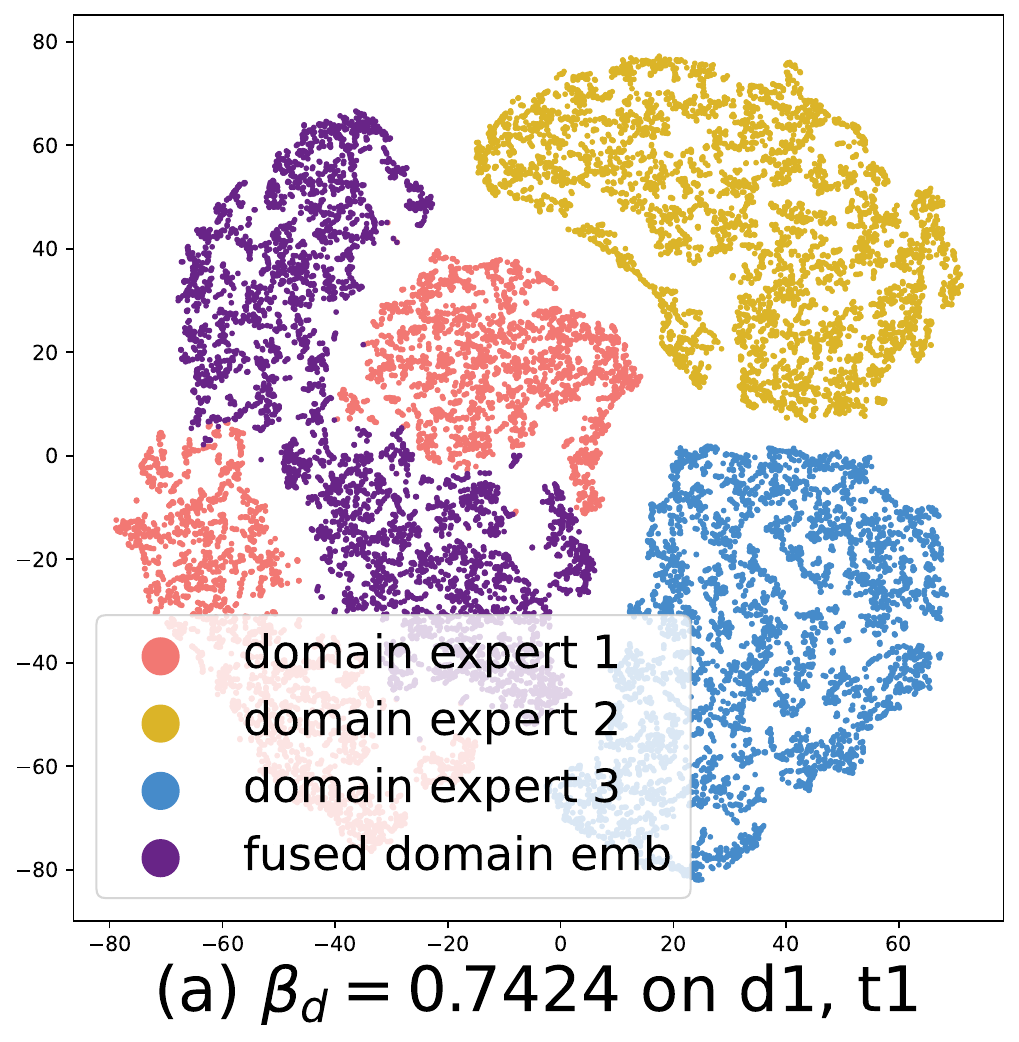}}}
    \vspace{-2mm}
{\subfigure{\includegraphics[width=0.49\linewidth]{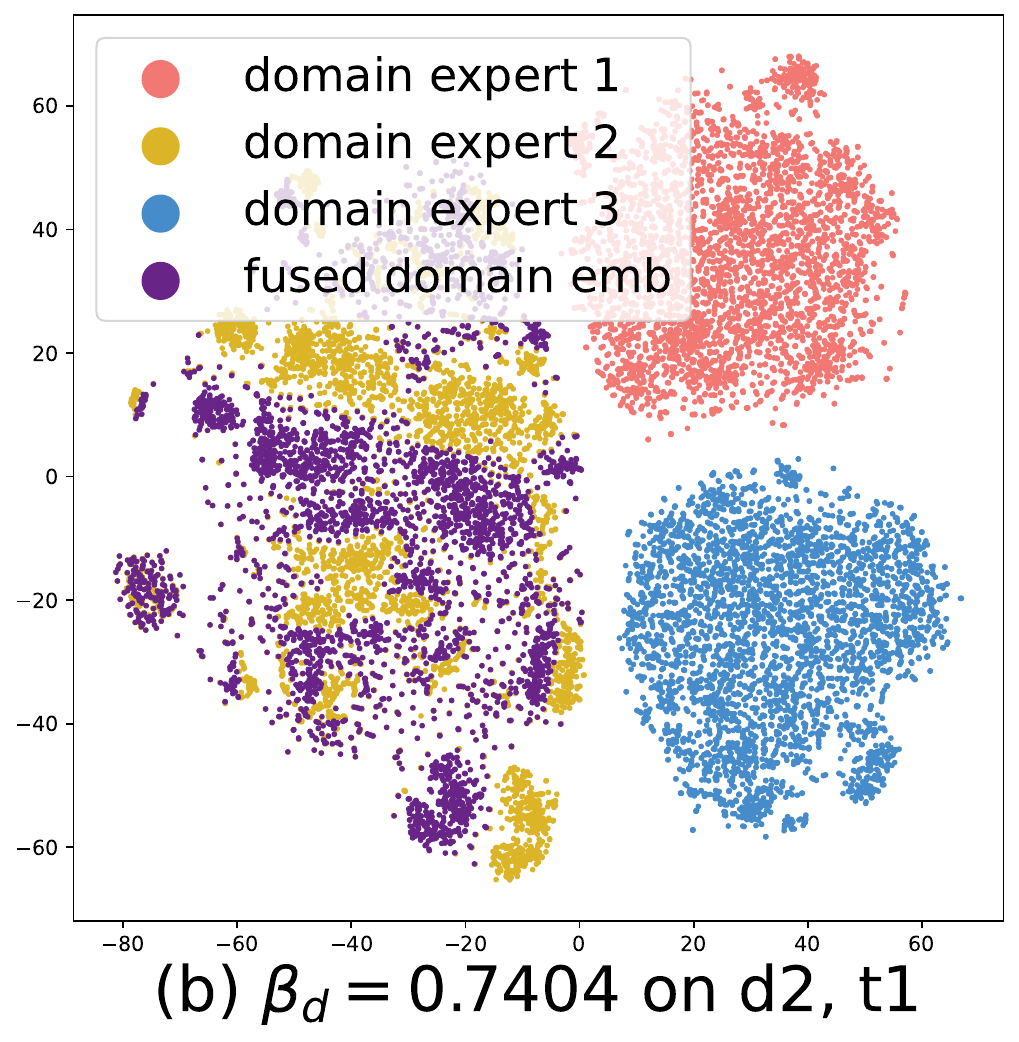}}}
    \vspace{-2mm}
{\subfigure{\includegraphics[width=0.49\linewidth]{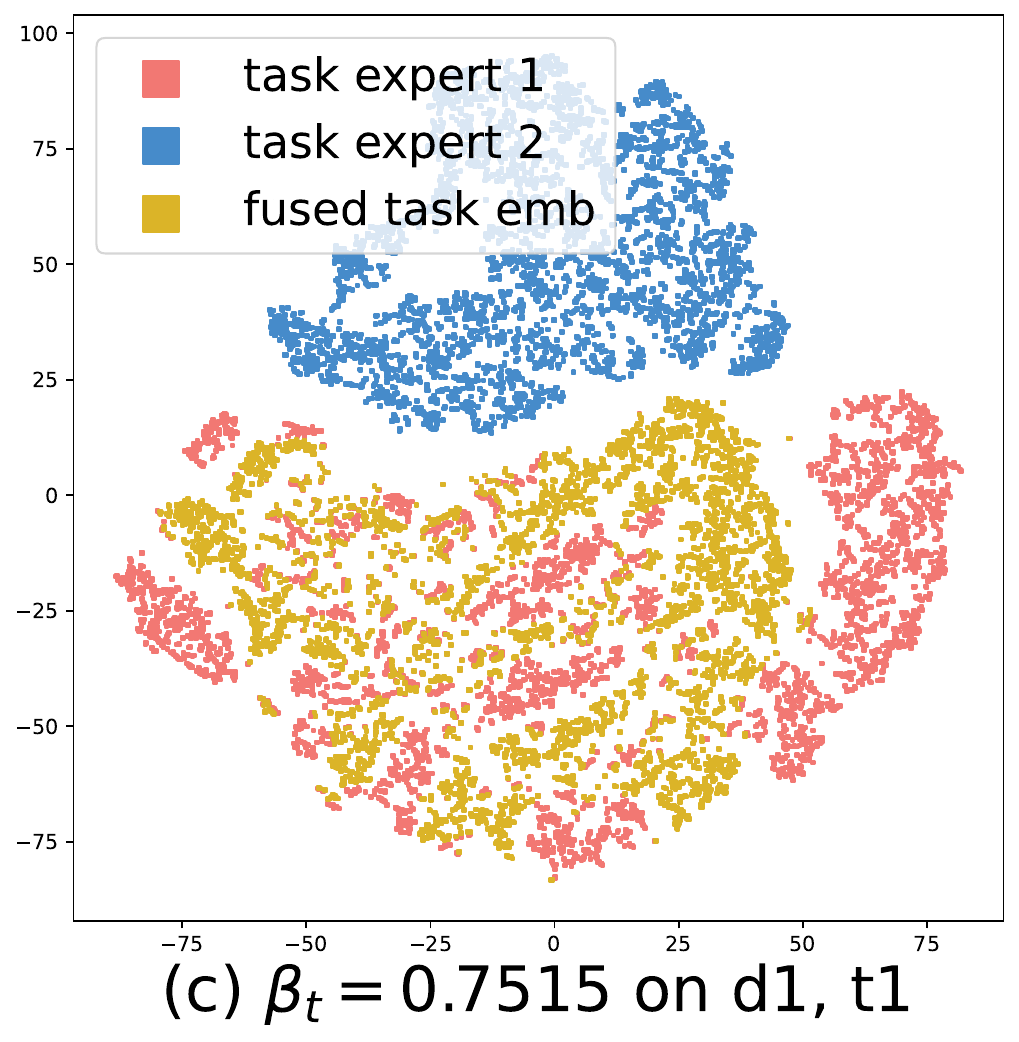}}}
    \vspace{-2mm}
{\subfigure{\includegraphics[width=0.49\linewidth]{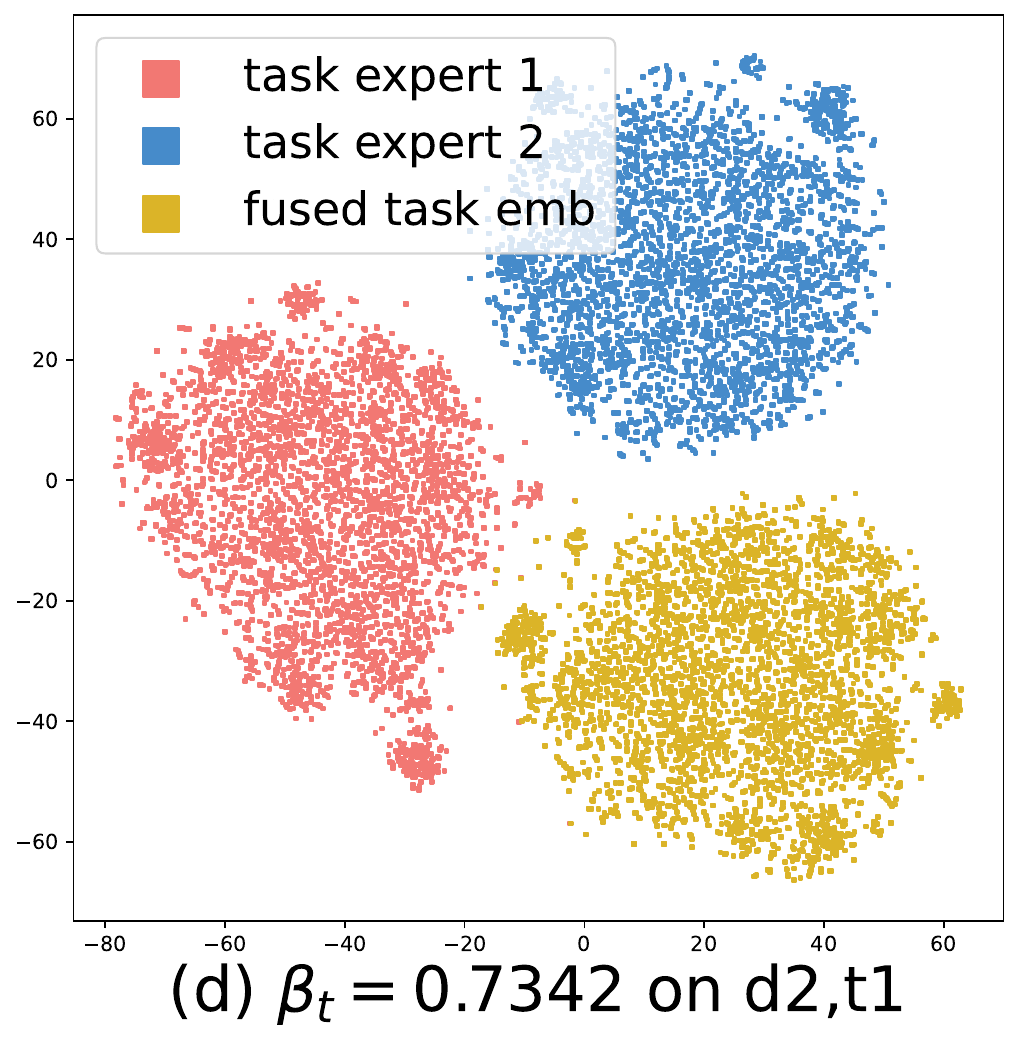}}}
    \vspace{-2mm}
{\subfigure{\includegraphics[width=0.49\linewidth]{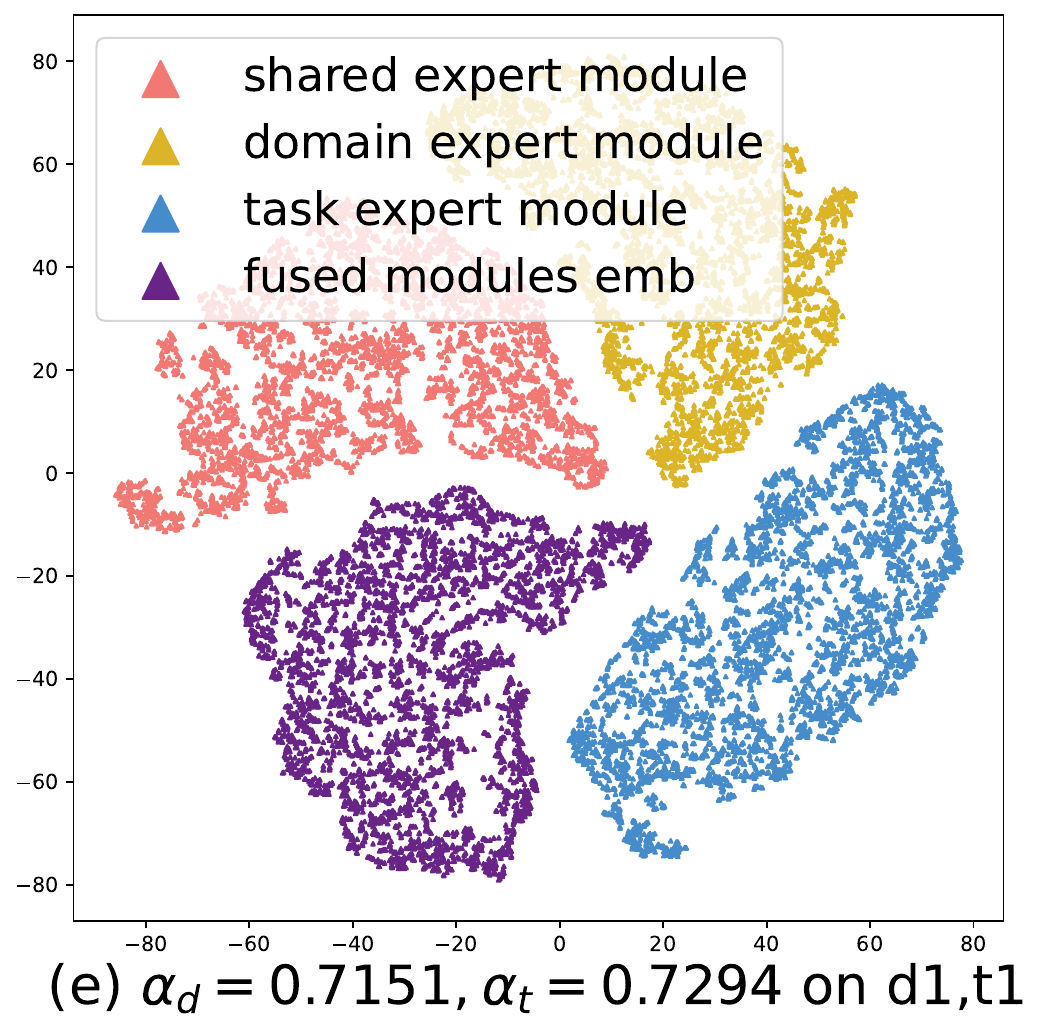}}}
    \vspace{-2mm}
{\subfigure{\includegraphics[width=0.49\linewidth]{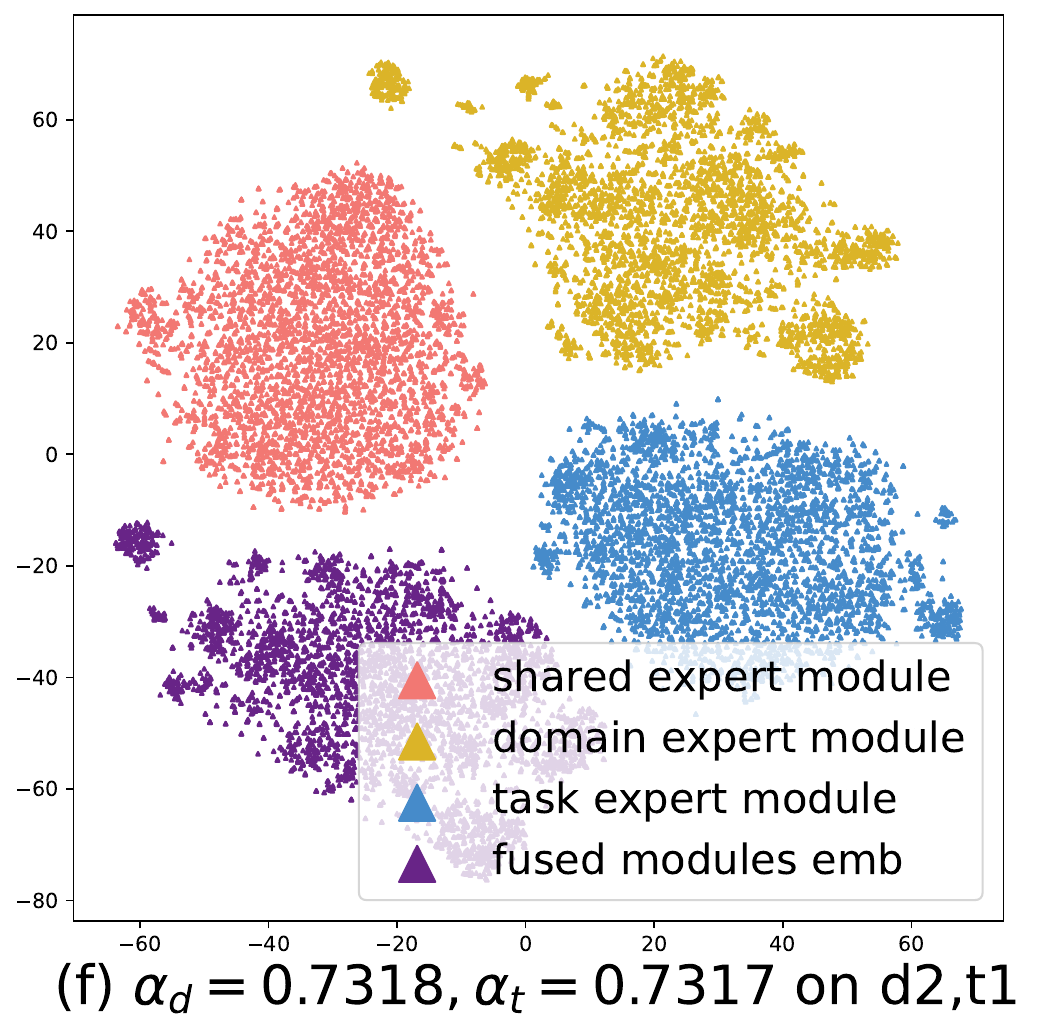}}}
    \vspace{-2mm}
    \caption{T-SNE results on MovieLens domain 1 task 1 (left column) and KuaiRand-Pure domain 2 task 1 (right column). }
    \label{fig:vis}
    \vspace{-3mm}
\end{figure}

\label{sec:visualization}



To provide an overview of the effectiveness of \name in multi-domain multi-task learning, we visualize the disentangled embeddings and fused embeddings. 
Figure \ref{fig:vis} (a) and (b) show the embeddings learned by the domain expert module on MovieLens and KuaiRand-Pure, respectively, so do (c) and (d) by the task expert module and (e) and (f) by the multi-view expert learning layer.
Both domain expert modules yield fused domain embeddings with a similar distribution with domain embedding (domain 1 in (a) and domain 2 in (b)), which means the domain-specific expert plays the dominant role in the knowledge fusion.
In terms of the task expert module, domain 2 of KuaiRand-Pure attains fused embedding with optimal $\beta_t$ independent of its components.
This indicates different task experts hold different views on the same domain embedding, and the model attains a balanced distribution among multiple experts, which accords with our expectations.
Similar phenomena occur on the fused modules embedding of both datasets, where none of the module outputs can replace the fused embedding.



\subsection{Ablation Study}

\begin{table}[t!]
\renewcommand{\arraystretch}{1.2}
\centering
\setlength{\abovecaptionskip}{1mm}
\setlength{\belowcaptionskip}{-4mm}
\caption{Components analysis with averaged AUC on all domains and tasks. Results are the average of 5 individual runs. 
}
\label{tab:ablation}
\scalebox{0.94}{
\begin{tabular}{lcc}
\toprule
\textbf{Dataset} & MovieLens & KuaiRand-Pure\\
\hline
\textit{w/o} AutoML & {76.37} & {65.41}\\
Concat modules & 76.89& \underline{66.08}\\
Fully gated modules & \underline{76.92}& 65.87\\
\textit{w/o} domain module & 76.89& 65.86\\
\textit{w/o} task module & 76.85& 65.92\\
\textit{w/o} domain\&task module & 76.80 & 65.80\\
\textbf{\name} & \textbf{77.02} & \textbf{66.37}\\
\bottomrule
\end{tabular}}
\vspace{-2mm}
\end{table}

To further investigate the effectiveness of \name, we conduct ablation studies by removing key components individually. 
Specifically, we present the results of the following variants:

\begin{itemize}[leftmargin=*]
\item \textbf{w/o AutoML.}
To approximate the manually designed fusion weights, we set the two-level fusion weights $\alpha_d$, $\alpha_t$, $\beta_d$, and $\beta_t$ as constraints by fixing them at 0.5.
\end{itemize}

\begin{itemize}[leftmargin=*]
\item \textbf{Concat modules.}
We concatenate the three modules outputs $\boldsymbol{\mathcal{S}}({\boldsymbol{h}_d})$, $\boldsymbol{\mathcal{D}}({\boldsymbol{h}_d})$, and $\boldsymbol{\mathcal{T}}({\boldsymbol{h}_d})$, rather than adding them as \name.
\end{itemize}

\begin{itemize}[leftmargin=*]
\item \textbf{Fully gated modules.}
In order to explore an alternative fusion mechanism, we evaluate the performance of fully gating the outputs of all modules instead of controlling them with weights.
\end{itemize}

\begin{itemize}[leftmargin=*]
\item \textbf{w/o domain module.}
We omit the domain expert module.
\end{itemize}

\begin{itemize}[leftmargin=*]
\item \textbf{w/o task module.}
We omit the task expert module.
\end{itemize}

\begin{itemize}[leftmargin=*]
\item \textbf{w/o domain\&task module.}
We omit both domain and task expert modules.

\end{itemize}

Based on the results presented in Table \ref{tab:ablation}, several conclusions can be drawn:
\textbf{\textit{(1)}} The adaptive fusion weights optimized by AutoML significantly improve performance, demonstrating their contribution to model flexibility and generality.
\textbf{\textit{(2)}} Concatenating the modules' outputs together yields inferior results compared to directly adding them. The latter approach maintains the modules' outputs in the same embedding space, enabling effective information transfer across domains and tasks.
\textbf{\textit{(3)}} The gated module fusion achieves moderate results. We claim that disentangling these modules into different factors and employing an effective fusion mechanism enhances multi-domain multi-task recommendation.
It is more effective than fusing by a unified gate.
\textbf{\textit{(4)}} We observe that the importance of domain modules and task modules differs in the two datasets, but removing these two modules clearly hurts performance.
Therefore, both modules contribute to performance, and different data depend on both information components in different ways, which emphasizes the importance of our adaptive solution.

\subsection{Hyper-Parameter Analysis}
In this subsection, we test several key hyper-parameters of \name to explore their impacts on the performance.
Specifically, we tune the learning rate of model parameters from $\{1e-4, 3e-4, 1e-3, 3e-3, 1e-2\}$, and present in Figure \ref{fig:hyper} (a).
From the results, we can observe that a relatively larger learning rate prompts the performance in general. Different datasets achieve the best results with different learning rates, with MovieLens peaking at $1e-2$ and KuaiRand-Pure peaking at $3e-3$.

Besides, we test the number of expert networks $N$ in the shared expert module $\boldsymbol{\mathcal{S}}$ from $\{1, 2, 3, 4, 5, 6\}$, and illustrates in Figure \ref{fig:hyper} (b).
According to the comparison, we can find that the two datasets reach the optimal result with different $N$, \ie $N=1$ for MovieLens and $N=4$ for KuaiRand, which shows different multi-domain multi-task scenarios prefers the balance between common knowledge, domain-aspect knowledge, and task-aspect knowledge in different degrees.
Besides, more expert networks than the optimal value may cause overfitting problems, and performance decreases as $N$ grows.
\begin{figure}[!t]
\setlength{\abovecaptionskip}{-1mm}
\setlength{\belowcaptionskip}{-4mm}
{\subfigure{\includegraphics[width=0.9\linewidth]{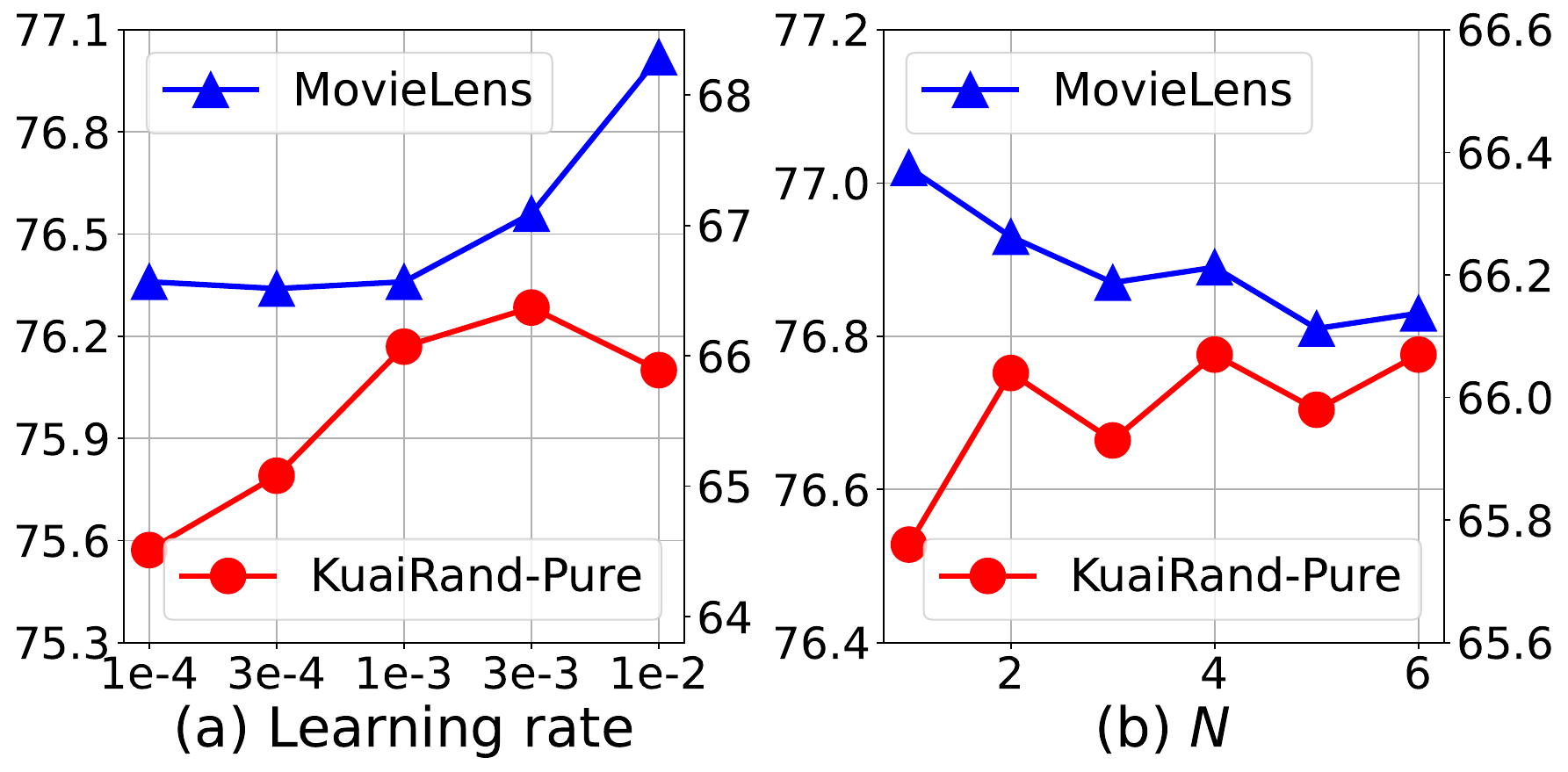}}}
%
    \caption{Hyper-parameter analysis results.
    }
    \label{fig:hyper}
\end{figure}

\vspace{-3mm}
\section{Related Works}
This section briefly reviews the representative methods in multi-domain recommendation, multi-task recommendation, and multi-domain multi-task recommendation.

\vspace{-2mm}
\subsection{Multi-Domain Recommendation}
MDR methods have gained research attention for their ability to transfer cross-domain knowledge in diverse business scenarios.\cite{star,dredze2010multi,joshi2012multi,li2020improving}.
Generally, existing MDR methods can be categorized into three groups: cross-domain mapping \cite{ddtcdr,yan2019deepapf}, MTR-based \cite{mmoe,star,causalint}, and dynamic weight-based \cite{adasparse}.
As a typical cross-domain mapping-based method, DDTCDR \cite{ddtcdr} utilizes orthogonal mapping for multi-domain embedding transfer so that the auxiliary information can be incorporated.
The key idea of MTR-based methods is to treat the multi-domain task as the multiple objectives optimization problem~\cite{mmoe,ple}. 
For example, CausalInt \cite{causalint} utilizes disentangled representation learning to capture scenario-invariant information, and incorporates a TransNet to use information from different scenarios.
Dynamic weight methods aim to generalize domain-specific knowledge by adjusting the weights of domain-specific modules, \eg AdaSparse \cite{adasparse} introduces adaptively sparse structures with domain-aware neuron-level weighting factors to identify and prune redundant neurons.
Unlike MDR methods, our \name leverages the advantages of multi-objective optimization to simultaneously capture multi-domain and multi-task aspects.


\subsection{Multi-Task Recommendation}
MDR frameworks \cite{anbomtl,mmoe,ple,forkmerg} aim to improve the performance of individual tasks by utilizing common dependencies among tasks.
They can be categorized into architecture-based \cite{adatt,mmoe} and optimization-based approaches \cite{forkmerg, adatask}.
SharedBottom \cite{shared_bottom} is one of the earliest architecture-based methods, which shares the hard structure across tasks.
Expert network sharing is another popular paradigm, \eg MMoE \cite{mmoe} uses fully shared experts to capture information across multiple tasks. PLE \cite{ple} incorporates task-specific experts to handle task-specific information, and AdaTT \cite{adatt} introduces a branch for linearly combining task-specific experts to model tasks jointly.
For the optimization-based methods, they primarily address specific challenges, including negative transfer and multi-objective trade-off \cite{mtl_balance1,mtl_balance2,mtl_survey}.
To tackle the negative transfer problem, MetaBalance \cite{metabalance} introduces a relaxation factor to balance the gradient magnitude proximity between auxiliary and target tasks. 
AdaTask \cite{adatask} addresses gradient conflicts among training objectives by employing a task-specific optimization strategy.
ForeMerge \cite{forkmerg} proposes balancing the auxiliary task loss weights to alleviate the negative transfer among tasks.
In terms of multi-objective trade-off issues, 
MTA-F \cite{mtl_balance1} explores the trade-off between group fairness and accuracy in multi-task learning to capture the multi-dimensional Pareto frontier.
Wang \etal \cite{mtl_balance2} address the trade-off between minimizing task training conflicts and enhancing model's multi-task generalization ability.
Existing MTR methods such as MMoE \cite{mmoe} and PLE \cite{ple} can be the special cases of our method, which considers both task-specific and common expert networks.

\subsection{Multi-Domain Multi-Task Recommendation}
Recently, there have been efforts towards multi-domain and multi-task recommendations \cite{m2m, pepnet, m3rec}, aiming to leverage the benefits of both domain and task simultaneously.
M2M \cite{m2m} employs a meta-unit to capture inter-scenario correlations and scale to new scenarios. It incorporates a meta-attention module for diverse inter-scenario correlations and a meta-tower module to enhance scenario-specific features.
PEPNet \cite{pepnet} leverages personalized prior information and dynamically scales units using gate mechanisms. This enables personalized selection and modification for users across multiple domains and tasks.
M3Rec \cite{m3rec} designs a Meta-Item-Embedding Generator (MIEG) and a User-Preference Transformer (UPT) to unify the representation of users and items, which can capture non-i.i.d behaviors across scenarios.
Existing methods usually rely on complex architectures, which leads to limited generality. 
In contrast, our framework employs disentangled modules and a self-adaptive two-level fusion mechanism, and optimizes based on data and task characteristics
, which is a promising and versatile solution.


\section{Conclusion}

Multi-domain recommendation and multi-task recommendation have made great progress in assisting knowledge fusion and improving recommendation accuracy.
However, cross-domain and cross-task knowledge transfer is more important in practical recommendation and cannot be handled well by existing methods.
In this paper, we first identify the long-neglected cross-domain and cross-task seesaw problem.
Then, we propose a framework \name to adaptively address MDMT recommendation for the first time.
It aims to solve general cross-domain and cross-task knowledge transfer through effective disentanglement and fusion mechanisms.
Extensive experiments on two public datasets demonstrate its outstanding efficacy in solving the MDMT seesaw problem.

\begin{acks}
This research was partially supported by Kuaishou, Research Impact Fund (No.R1015-23), APRC - CityU New Research Initiatives (No.9610565, Start-up Grant for New Faculty of CityU), CityU - HKIDS Early Career Research Grant (No.9360163), Hong Kong ITC Innovation and Technology Fund Midstream Research Programme for Universities Project (No.ITS/034/22MS), Hong Kong Environmental and Conservation Fund (No. 88/2022), and SIRG - CityU Strategic Interdisciplinary Research Grant (No.7020046, No.7020074).
Hongwei Zhao is funded by the Provincial Science and Technology Innovation Special Fund Project of Jilin Province, grant number 20190302026GX, Natural Science Foundation of Jilin Province, grant number 20200201037JC, and the Fundamental Research Funds for the Central Universities, JLU.
\end{acks}

\bibliographystyle{ACM-Reference-Format}
\balance
\bibliography{9Reference}

\end{document}